\documentclass[aps,preprint]{revtex4}%
\usepackage{amsfonts}
\usepackage{amsmath}
\usepackage{amssymb}
\usepackage{graphicx}%
\providecommand{\U}[1]{\protect\rule{.1in}{.1in}}

\begin{document}
\title{Comment on "Chiral tunnelling and the Klein paradox in graphene"}
\author{M. Ahsan Zeb}
\email{maz24@cam.ac.uk}
\affiliation{Department of Earth Sciences, University of Cambridge, UK.}
\date{12 Oct, 2010}
\begin{abstract}
Arising from the Article: Nature Phys. 2, 620-625 (2006), By M. I. Katsnelson,  K. S. Novoselov, and  A. K. Geim. 
\end{abstract}
\maketitle

To the Editor-- The authors of ref.\cite{c1} consider transmission through a square potential barrier of chiral particles in mono and bilayer graphene and non-chiral particles in a zero gap semiconductor with parabolic (and symmetric) electron and hole bands. They compare the transmission probabilities in the three cases and attribute the differences to the different chiralities involved. 

Although, the authors do not provide explicit details of their calculations in case of tunneling of non-chiral particles, the expression for the transmission amplitude they present indicates that they only consider the band having the propagating states, i.e., the band where energy of the incident particles lies. In other words, at positive (negative) energies only electron (hole) states have been considered while the evanescent states belonging to the hole (electron) band have been neglected. 

Simultaneously considering both bands leads to entirely different results: Just like the case of chiral particles in bilayer graphene (see Fig.3 in ref.\cite{c1}), the transmission probability of normally incident non-chiral particles decays with the barrier width. However, there is a clear difference between the angular dependence of the transmission probabilities of these chiral and non-chiral particles. Indeed, for the latter, the transmission probability always peaks at normal incidence and there are no resonances. 

The hypothetical semiconductor considered above, a system that has the same band structure as that of bilayer graphene but whose particle states do not possess any kind of pseudospin chirality, can be described by a Hamiltonian that has the following matrix elements in some basis. $H_{11}=H_{22}=0$ and $H_{12}=\frac{-\hbar^{2}}{2m}(\partial_{x}^{2}+\partial_{y}^{2})$, where $\partial_{x,y}$ are the differential operators and m is the effective mass of the particles taken to be the same as that of chiral particles in bilayer graphene. It is unclear whether such a system can be realised due to the constraints on the relative strengths of various hopping terms involved. Nevertheless, we can compare this system to bilayer graphene to find out about the effects of chirality on the properties of the latter. In the comparison made above, we find that chirality leads to anisotropy but the decaying feature, in the plot of transmission probability of normally incident particles against the barrier width, can only be attributed to the band structure of bilayer graphene.

\begin{acknowledgments}
I am thankful to Nicholas Bristowe and Emilio Artacho for their help and advice.
\end{acknowledgments}

\end{document}